\documentclass[11pt,twocolumn,oneside]{article}
\usepackage[a4paper,top=3cm,bottom=3cm,left=1.7cm,right=1.7cm]{geometry}
\usepackage[babel]{csquotes}
\usepackage{graphicx}
\usepackage{subfigure}
\usepackage{amsmath,amssymb}
\usepackage[colorlinks,citecolor=blue,linkcolor=blue,urlcolor=blue]{hyperref}
\usepackage{mathrsfs}
\usepackage{authblk} 
\usepackage{abstract}
\usepackage{titlesec} 
\usepackage[T1]{fontenc}
\usepackage[margin=4pt, font=footnotesize, labelfont=bf, labelsep=period]{caption}	[2004/07/16]
\usepackage{mathtools}	
\usepackage{cuted}
\usepackage{flushend}
\usepackage{balance}

\renewcommand\thesection{\Roman{section}}

\titleformat{\section}{\large\scshape\bfseries\centering}{\thesection.}{.7em}{}
\titleformat{\subsection}{\scshape\bfseries}{\thesubsection.}{.7em}{}

\def\be{\begin{equation}}
\def\ee{\end{equation}}

\newcommand{\AppSep}{%
\begin{strip}%
\vspace{-4ex}%
\center
\rule{4cm}{.5pt}
\vspace{1ex}%
\end{strip}}

\newcommand{\f}{\frac}
\newcommand{\tl}{\tilde}

\def\p{\partial}

\newcommand{\R}{{\mathbb R}}

\newcommand{\Ref}[1]{(\ref{#1})}

\renewcommand{\a}{\alpha} \renewcommand{\b}{\beta} \newcommand{\g}{\gamma}
\renewcommand{\d}{\delta}  \newcommand{\eps}{\epsilon}  
 \renewcommand{\th}{\theta}      \renewcommand{\l}{\lambda}
\let\m=\mu  \let\n=\nu  \let\r=\rho \newcommand{\s}{\sigma}       \let\om=\omega
 \let\D=\Delta   \let\L=\Lambda \let\Si=\Sigma \let\Om=\Omega

\title{\Large{\textbf{\textsc{Non-singular rotating black hole with a time delay in the center}}}}
\author[1]{Tommaso De Lorenzo%\thanks{tommaso.de-lorenzo@cpt.univ-mrs.fr}
}
\author[1,2]{Andrea Giusti
}
\author[1]{Simone Speziale%\thanks{simone.speziale@cpt.univ-mrs.fr}
}

\affil[1]{\textit{Centre de Physique Theorique, CNRS-UMR 7332, Aix-Marseille Universit\'e \& Universit\'e de Toulon, Case 907, Campus de Luminy, 13288 Marseille, France
\newline 
$^2$\textit{Dip. di Fisica e Astronomia, Universit\`a di Bologna, via Irnerio 46, I-40126 Bologna, Italy}
\affil[2]{\textit{Dipartimento di Fisica e Astronomia, Universitˆ di Bologna via Irnerio 46, I-40126 Bologna, Italy}}}
\vspace{-2.5ex}
}
\date{\small(Dated: 29 October 2015; v2: 2 February 2016)\vspace{-4ex}}

\begin{document}
\twocolumn[
\vspace{-7ex}\maketitle%
\begin{list}{}{\leftmargin=3em\rightmargin=\leftmargin}\item\relax
\small\textbf{\textsc{Abstract.}} 
As proposed by Bambi and Modesto, rotating non-singular black holes can be constructed via the Newman-Janis algorithm. Here we show that if one starts with a modified Hayward black hole with a time delay in the centre, the algorithm succeeds in producing a rotating metric, but curvature divergences reappear. To preserve finiteness, the time delay must be introduced directly at the level of the non-singular rotating metric. This is possible thanks to the deformation of the inner stationarity limit surface caused by the regularisation, and in more than one way. We outline three different possibilities, distinguished by the angular velocity of the event horizon. Along the way, we provide additional results on the Bambi-Modesto rotating Hayward metric, such as the structure of the regularisation occurring at the centre, the behaviour of the quantum gravity scale alike an electric charge in decreasing the angular momentum of the extremal black hole configuration, or details on the deformation of the ergosphere.
\end{list}\par\vspace{3ex}%
]
\saythanks
%-----------------------------------------------------
\section{Introduction}
%-----------------------------------------------------

Non-singular black holes provide an interesting case study for possible low-energy effects of quantum gravity. The main idea is that quantum mechanical effects remove the central singularity, while the observables large scale features such as event horizons and orbiting potential profiles are preserved. This scenario, appealing by itself as it may cure the fact that general relativity predicts its own downfall, could also provide a simple resolution to the information-loss paradox (see e.g. \cite{hayward2006formation,Frolov:BHclosed,hossenfelder2010conservative}), and lead to observational consequences (see e.g. \cite{Barrau:2014hda,Barrau:2014yka,Barrau:2015uca}.) 

Non-singular black holes have a long history (e.g. \cite{hayward2006formation,Frolov:BHclosed,bardeen1968non, Frolov:1981mz, Roman:1983zza,Casadio:1998yr, Mazur:2001fv, dymnikova2002cosmological,Visser:2009pw,Falls:2010he,Modesto:2010rv,BambiModestoKerr,Bambi:2013caa,Haggard:2014rza,Mersini-Houghton:2014yq,DeLorenzo:2014pta}), and appear in different contexts such as non-linear electrodynamics \cite{AyonBeato:2000zs}, asymptotic safety \cite{LitimASBH,Saueressig:2015xua}, non-local modified theories of gravity \cite{ModestoNonLocalBH,Frolov:2015bta}, and loop quantum gravity \cite{ModestoLQGBH,AshtekarBojowald}, where a scenario analogue to the cosmological bounce \cite{Ashtekar:2006rx} leads to the idea of Planck star \cite{Rovelli:2014cta}. The motivations for the present paper are to extend our knowledge of these systems and getting them closer to physical applications, by studying a model of a rotating black hole with a time delay in its regular centre, a simple effect to be expected by analogy with internal stellar structures.

As it turns out, most models in the literature describe spherically symmetric objects. The simplicity of the metric helps keeping things under control in a domain where the classical energy conditions are not satisfied, and anything and everything can a priori happen.
However, rotation must be included if one is willing to model physical objects. A well-known procedure to introduce rotation in general relativity is the Newman-Janis (NJ) algorithm \cite{NewmanJanis65}. In a seminal paper \cite{Caravelli:2010ff}, the algorithm was applied to Loop Quantum Gravity inspired non-singular black holes, while in \cite{BambiModestoKerr}, Bambi and Modesto applied it to both the Hayward \cite{hayward2006formation} and the Bardeen \cite{bardeen1968non} metrics, two of the most popular models of non-singular black holes. Their results show that the algorithm succeeds in generating a rotating metric, with an event horizon and an ergosphere, while preserving a non-singular centre. 
In Section II, we first review the Bambi-Modesto model and provide a few more details on its structure, focusing on the case of the rotating Hayward metric. In particular, we discuss how the quantum gravity correction approaches the two horizons and decreases the value of the angular momentum of the extremal black hole, in this sense behaving as an electric charge. Secondly, we show that the ergosphere is deformed in a way as to increase the velocity rotation thus decrease maximal energy extraction, and that the inner stationarity limit surface acquires an hour-glass shape, with the consequence that the outer time-like Killing vector is time-like also in the regular centre. Similar effects have been reported in \cite{Ghosh:2015pra} for the rotating Bardeen metric.
Of course, being regular the metric could make sense also beyond the extremal case, in which there are no more horizons, and the two stationary limit surfaces merge defining a compact surface with the topology of a 3-torus. 

In Section III, we apply the same algorithm to the modified Hayward metric proposed in \cite{DeLorenzo:2014pta}, which features a non-trivial time delay between an asymptotic observer and an observer in the centre, as well as a matching to the expected $1/r^3$ fall off from the 1-loop corrections \cite{DonoghueKerr}.
We find that the algorithm succeeds in generating a rotating metric, however it reintroduces a singularity in the centre, for no matter how small the time delay is. Hence, absence of physical singularities is not a property preserved by the algorithm in general applications, and a rotating metric with time delay can not be derived in this way. In Section IV we show how this difficulty can be circumvented introducing the time delay directly in the Bambi-Modesto rotating Hayward metric. This is possible thanks to the deformation of the inner stationarity limit surface caused by the regularisation. The procedure is however not unique, and we discuss three possible cases, which have the same time delay, and are distinguished by the area and angular velocity of the horizon. For the resulting metrics, the Kretschmann curvature invariant is finite at the centre, but discontinuous, like the original rotating Hayward metric. Indeed, although there is some freedom in the time-delay function, we find that it can never make the  Kretschmann invariant  continuous. The invariant also shows a bizarre non-monotonic behaviour with a peak near the inner horizon. The value of the peak depends also on the time delay, thus requiring that the curvature be sub-Planckian everywhere gives a bound on the maximal time delay. The metrics also violate the weak energy conditions, in both the negativity of energy density and the magnitude of pressures, but only in small regions around the inner horizon.

Throughout the paper, calculations of curvature invariants, their plots and series expansions are performed with Wolfram's Mathematica and the xTensor package.

%-----------------------------------------------------
\section{Non-singular rotating black holes}
%-----------------------------------------------------

In '65 \cite{NewmanJanis65} Newman and Janis observed from the properties of Weyl curvature and null geodesics  congruences, that the physical differences between the Schwarzschild and Kerr metrics are captured by a shift $r\mapsto r+i a \cos\th$, where $J=aM$ is the angular momentum of the hole.  This led to a famous algorithm that carries their name and that maps some solutions to Einstein's equations to new solutions with non-vanishing angular momentum.
The details of the algorithm are well covered in the literature (e.g. \cite{AdamoNewman14,Drake:1998gf}) and we do not review them here, but merely state the main result. For a static seed metric with spherical symmetry in advanced null coordinates,
\be\label{gensph}
ds^2 = - f(r)du^2 - 2\sqrt{f(r) h(r)} du dr + r^2 d^2\Om,
\ee
the algorithm gives a non-diagonal metric, given by
\begin{equation}\label{NJg}
\begin{split}
g_{uu} &= -\tl{f}, \quad g_{u r} = - \sqrt{\tl f \tl h}, \quad g_{\theta \theta} = \Sigma,\\
g_{u \varphi} &= a \sin ^2 \theta \left(\tilde{f} - \sqrt{\tl f \tl h} \right), \\
g_{r \varphi} &= a \sin ^2 \th \sqrt{\tl f \tl h},  \\
g_{\varphi \varphi} &= \sin ^2 \th \left[ \Si + a^2 \sin ^2 \th \left( 2 \sqrt{\tl f \tl h} - \tilde{f}  \right)  \right].
\end{split}
\end{equation}
Here 
\be\label{defS}
\Sigma = r^2 + a^2 \cos^2\th,
\ee
and the tilde over functions means that the following substitution has taken place,
\be\label{NJ1}
\f1r \mapsto \f r\Sigma,
\qquad r^2 \mapsto \Sigma.
\ee
The procedure, originally applied to the Reissner-Nordstrom solution to obtain the Kerr-Newman one, has later been shown to apply to more general metrics of the Kerr-Schild form (albeit not to all of them), and successfully used to generate new solutions, see e.g. \cite{Stephani:2003tm, AdamoNewman14} for reviews.

The idea of Bambi and Modesto \cite{BambiModestoKerr} is to apply the algorithm to obtain rotating metrics starting from existing models of spherically symmetric non-singular black holes. To do so, one needs to partially extend the algorithm: as one can see from \Ref{NJ1}, terms like $r$ and $r^2$ are transformed in a different way; in non-singular black holes metrics additional powers of $r$ appear, which require independent prescriptions. For instance, Hayward's black hole \cite{hayward2006formation} looks like the Schwarzschild metric, for which
\be
f(r) \equiv F(r):=1-\f{2m}r,
\ee 
where now the mass $m$ is replaced by an $r$-dependent function given by
\be\label{MHay}
M(r) = m \frac{r^3}{r^3 + 2 m L^2}\;.
\ee
The spacetime is asymptotically flat and $m$ still measures the ADM mass of the metric, while
 $L$ is a quantum gravity parameter of the order of the Planck scale. 
The metric obtained from the NJ algorithm has the form \Ref{NJg} with
\be
\tl f= 1/ \tl h \equiv  \tl F(r) = 1-\f{2 \tl M(r,\th) r}\Si\;,
\ee
and the prescription proposed by Bambi and Modesto in \cite{BambiModestoKerr} is
\be\label{NJ3}
\tl M(r,\th) = m \frac{r^{3+\g} \Si^{-\g/2}}{r^{3+\g} \Si^{-\g/2} + 2 m L^2 r^{\d} \Si^{-\d/2}}\;.
\ee
It amounts to splitting $r^3=r^2 r$ and allowing arbitrary mixtures of \Ref{NJ1}, namely
\be
r^3 \mapsto r^{3+\g} \Si^{-\g/2}, \qquad \g\in\R
\ee
and further taking into account the freedom that comes from the different transformation of $r$ and $1/r$ that can be deduced from \Ref{NJ1}. Bambi and Modesto applied a similar prescription also to the function $M(r)$ proposed by Bardeen \cite{bardeen1968non}, obtaining a second class of non-singular rotating black holes. In this paper we restrict attention to the Hayward case. For more details on the rotating Bardeen hole see \cite{Ghosh:2015pra}. Our main intent is to introduce in the metric a non-trivial time delay in the center, as already proposed for the non-rotating case in \cite{DeLorenzo:2014pta}, and further studied in \cite{Debnath:2015hea}.

The Bambi-Modesto rotating Hayward spacetime looks like Kerr, with an extra `regulating' factor induced by \Ref{MHay}.
A special case of interest which presents considerable semplications is $\g=\d$, for which $\tl M(r,\th)\equiv M(r)$. In this case, one can transform to Boyer-Lindquist coordinates, with a single off-diagonal entry in the metric,
\begin{align}\label{BM1}
ds^2 &= - \left(1-\f{2M(r) r}\Si\right) dt^2 
\\ & \quad - \f{4a M(r) r \sin^2{\th}}{\Sigma}dtd\phi+\f{\Sigma}{\Delta}dr^2+\Sigma d\th^2 \nonumber \\
& \quad +\sin^2\th \left(r^2+a^2+\f{2a^2 M(r) r \sin^2\th}{\Sigma}\right)d\phi^2, \nonumber
\end{align}
where
\be\label{deltagd}
\D  := r^2 - 2 M(r) r + a^2 = 0.
\ee
In general, it is the $\th$-dependence in $\tl M(r,\th)$ that prevents this step, see \cite{BambiModestoKerr} for details.

For any $\g$ and $\d$, the metric describes an asymptotically flat spacetime, with a time-like and a space-like Killing vectors, and associated conserved charges given by the mass $m$ and the angular momentum per unit of mass $a$. It has an outer and an inner horizon, and two stationary limit surfaces. This is all very similar to Kerr's metric, which is promising for astrophysical applications \cite{BambiTestingKerr11,BambiTestingBardeen14}. The most important difference is, however, that the metric is everywhere regular. The singularity theorems are avoided because the energy conditions are violated. In the rest of this section we look at this metric in some detail, reviewing and at places continuing the analysis of \cite{BambiModestoKerr}.

\begin{figure}[ht]
\centering
\includegraphics[width=0.45\textwidth]{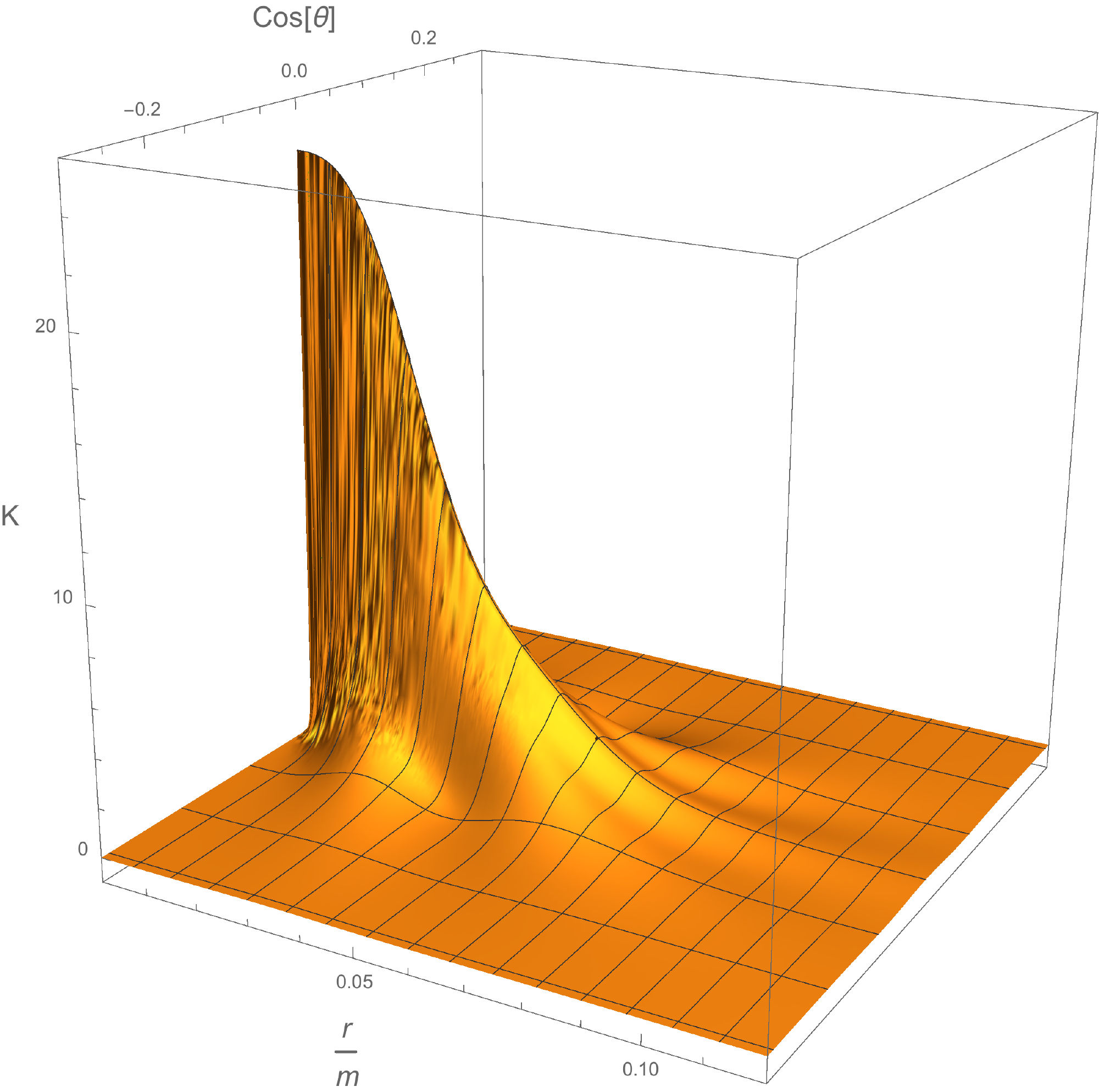}
\caption{The Kretschmann invariant ${\cal K}$ for the Bambi-Modesto metric in the case $\g=\d$ with $m=100$, $a=60$ and $L=1$ (in Planck units), with inner horizon located at $r_-/m \simeq 0.2$ and the event one at $r_+/m \simeq 1.8$. Notice the discontinuity at $r=0$, and the matching of the maximum value with the non-rotating case, $24/L^2$.}
\label{fig:K_CL}
\end{figure}

\paragraph{Regular but discontinuous center.} 
The Bambi-Modesto spacetime has no curvature singularities, as can be seen looking for instance at the Kretschmann invariant, ${\cal K}:=R^{\m\n\r\s}R_{\m\n\r\s}$. Its behaviour for $r\mapsto 0$ was given 
in \cite{BambiModestoKerr}, and it is finite as desired, but discontinuous: approaching $r=0$ along the equator, on finds ${\cal K}=24/L^4$, whereas approaching the centre away from the equator $\cal K$ vanishes identically.\footnote{The same discontinuity was previously observed in a model of regular rotating black hole proposed in \cite{Smailagic:2010nv}, motivated by non-commutative geometry. In that scenario, it is possible to remove both the singularity and the discontinuity with a rotating string of Planckian tension replacing the ring singularity.}
 A numerical plot, see Fig.~\ref{fig:K_CL},
confirms this discontinuity, and shows that the value in zero at the equator is the absolute maximum of $\cal K$. This is useful, because in order to make sense as semiclassical metrics, such non-singular black holes must have sub-Planckian curvature everywhere. Notice that the maximum is, quite surprisingly, the same as for the non-rotating Hayward metric, so the request of sub-Planckian curvature imposes the same lower bound on $L$ to a few units of the Planck length.

Let us comment some more on the discontinuity. The regular behaviour was referred to as a `de Sitter belt' in \cite{BambiModestoKerr}, and indeed, in the non-rotating Hayward case, the regularisation of the central singularity is achieved via a de Sitter behaviour near the origin. One may then expect that the ring singularity of the Kerr solution is regularised by a rotating de Sitter behaviour near the origin. However, this is not quite the case. First of all, when one applies the NJ algorithm to de Sitter, one does not obtain the conventional rotating de Sitter metric (defined setting $m=0$ in the Kerr-de Sitter metric of \cite{Gibbons:1977mu}), but rather a metric of type \Ref{NJg} which gives a non-trivial energy-momentum tensor, and thus does not solve the vacuum Einstein's equations with cosmological constant.\footnote{Indeed, this is a typical example in which the NJ algorithm does not work in its original meaning, in the sense that it produces a metric that does not solve the same Einstein's equations as the seed metric.}
Secondly, to make things more complicated, it turns out that the Bambi-Modesto metric does reproduce this NJ-de Sitter behaviour near the origin, but only at the equator. Away from the equatorial plane, the behaviour is completely different. This can be easily understood because of the dependence on $\g$ and $\d$ which is absent in the NJ-de Sitter case which only requires the basic prescription \Ref{NJ1}. But even in the case $\g=\d$, we have for instance $g_{uu}\simeq -1+r^4/(a^2\cos^2\th L^2)$ 
for the Bambi-Modesto metric, whereas the NJ-de Sitter has $g_{uu}= -1+\L \Si/3$, where $\Lambda$ is the cosmological constant.

We conclude from this discussion that the blind application of the NJ algorithm can introduce non-trivial structures which are hard to guess from the seed metric. The divergences we will find later on provide an additional example.

\begin{figure}[ht]
\centering
\includegraphics[width=0.45\textwidth]{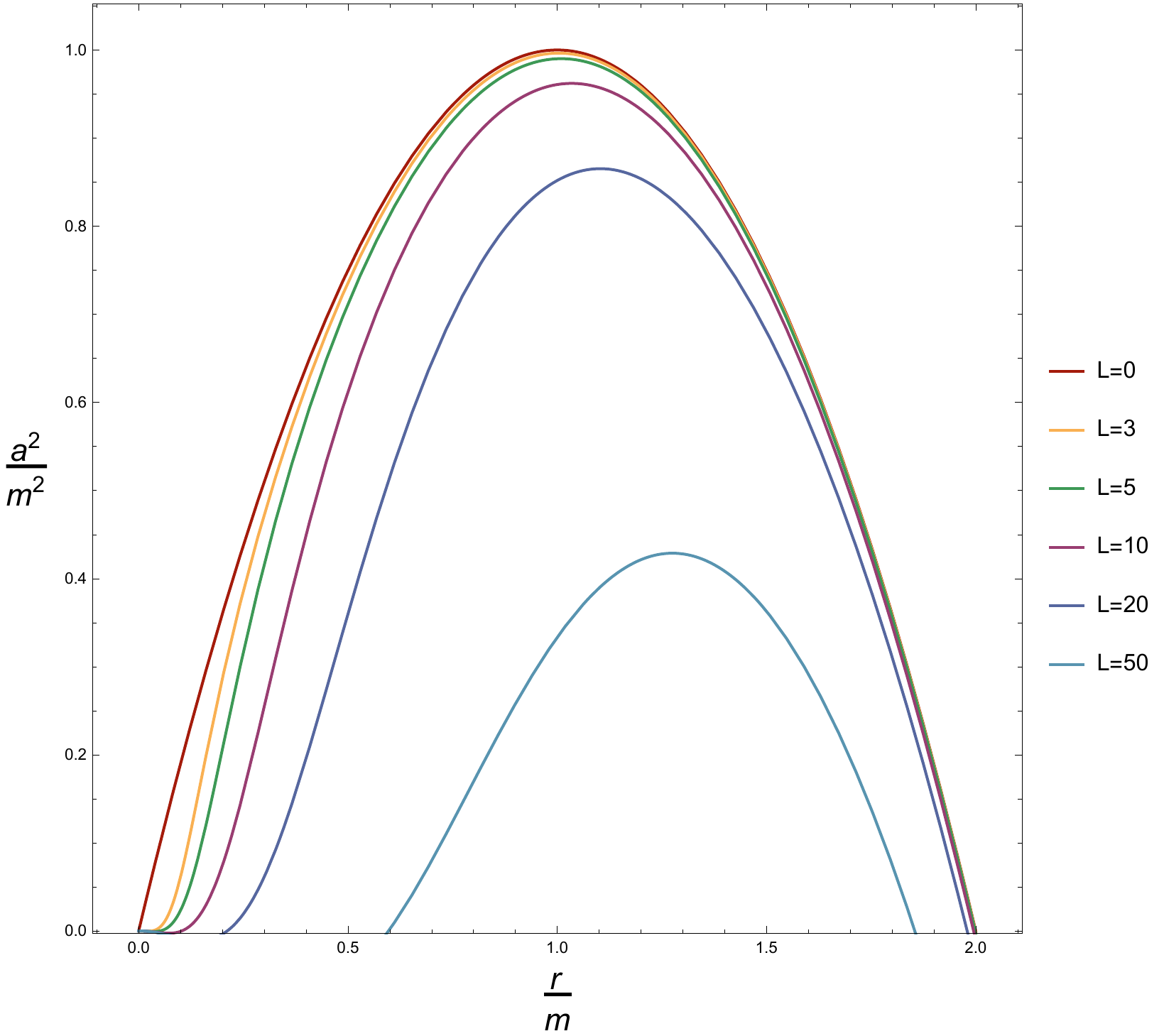}
\caption{The relation between horizons radius and the parameter $a/m$ for different values of the parameter $L$. 
Increasing $L$ makes the locations of the horizons converge and decreases the extremal value. A functional dependence $a_{\rm ex}(m,L)$ can be deduced interpolating the maxima of this plot.}
\label{FigH}
\end{figure}

\paragraph{Horizons.}
The Bambi-Modesto metric is basically Kerr with $m$ replaced by $M(r)$, so the positions of the horizons is given by the usual formula,
\be\label{delta}
\tl \D  := r^2 - 2 \tl{M}(r,\th) r + a^2 = 0.
\ee
The degree of this polynomial equation in $r$ is now dependent on the parameters $\g$ and $\d$. To give an explicit treatment, we focus on the case $\g=\d$, for which the $\th$-dependence drops and the equation is quintic. Numerical investigations show that for $L\ll m$ the equation has at most two real and positive solutions. As for Kerr, increasing the angular momentum makes the two horizons approach, until they merge for some extremal value $a_{\rm ex}(m,L)$. This is smaller than the Kerr value, $a_{\rm ex}=m$, and decreases with $L$, see Fig.~\ref{FigH}.
It thus has the same qualitative effect of the of adding an electric charge to a Kerr black hole. 

As in our motivations $L$ is close to the Planck scale, we can solve \Ref{delta} perturbatively in $L/m\ll 1$, and provide analytical details to the numerical results.
This gives the approximate locations
\be
r_\pm = r_\pm^{(0)} + \eps_\pm L^2 + o(L^3),
\ee
where 
\be
r_\pm^{(0)} = m\pm \sqrt{m^2-a^2}
\ee
are the usual locations of the horizons of a Kerr black hole, and where
\be\label{eps}
\eps_\pm = \f{2m^2 r^0_\pm}{-m r^0_\pm(4m^2- 3a^2)+ a^2(2m^2-a^2)}.
\ee
The outer and inner correction terms have different properties. It can be easily checked that $\eps_+$ is always negative for $a\leq m$, while $\eps_-$ is always positive. Hence, the two horizons approach each other, in agreement with the numerical analysis shown in Fig.~\ref{FigH}. Secondly, $|\eps_+|<1$ for $a\leq m$, so the approximated solution is valid no matter the value of the angular momentum. On the other hand, $\eps_-$ can become large in the slowly rotating limit, so the approximation breaks down. Indeed, for $a/m\ll 1$ we have
\begin{align}
& r_+ \simeq 2m -\f{L^2}{2m}, \\
& r_- \simeq \f{a^2}{2m}+L^2\left(\f{1}{2m}+\f{8m^3}{a^4}\right).
\end{align}
From the first approximation, we can also see that at first order,  the introduction of $L$ has the same effect of a an electric charge $Q=2L$.
Finally, the first order result can be used to compute the approximate area of the event horizon,
\begin{align}
A_H & = 16\pi m (r_+^{(0)}{})^2 \bigg[ 1-\f{L^2}{m (r_+^{(0)}{})^3 }  \\ \nonumber & \qquad  \times \Big(
2m^2 -\eps_+ [(6m^2-a^2) r_+^{(0)}-3m a^2]\Big) \bigg].
\end{align}

\begin{figure}[ht]
\centering
\includegraphics[width=0.45\textwidth]{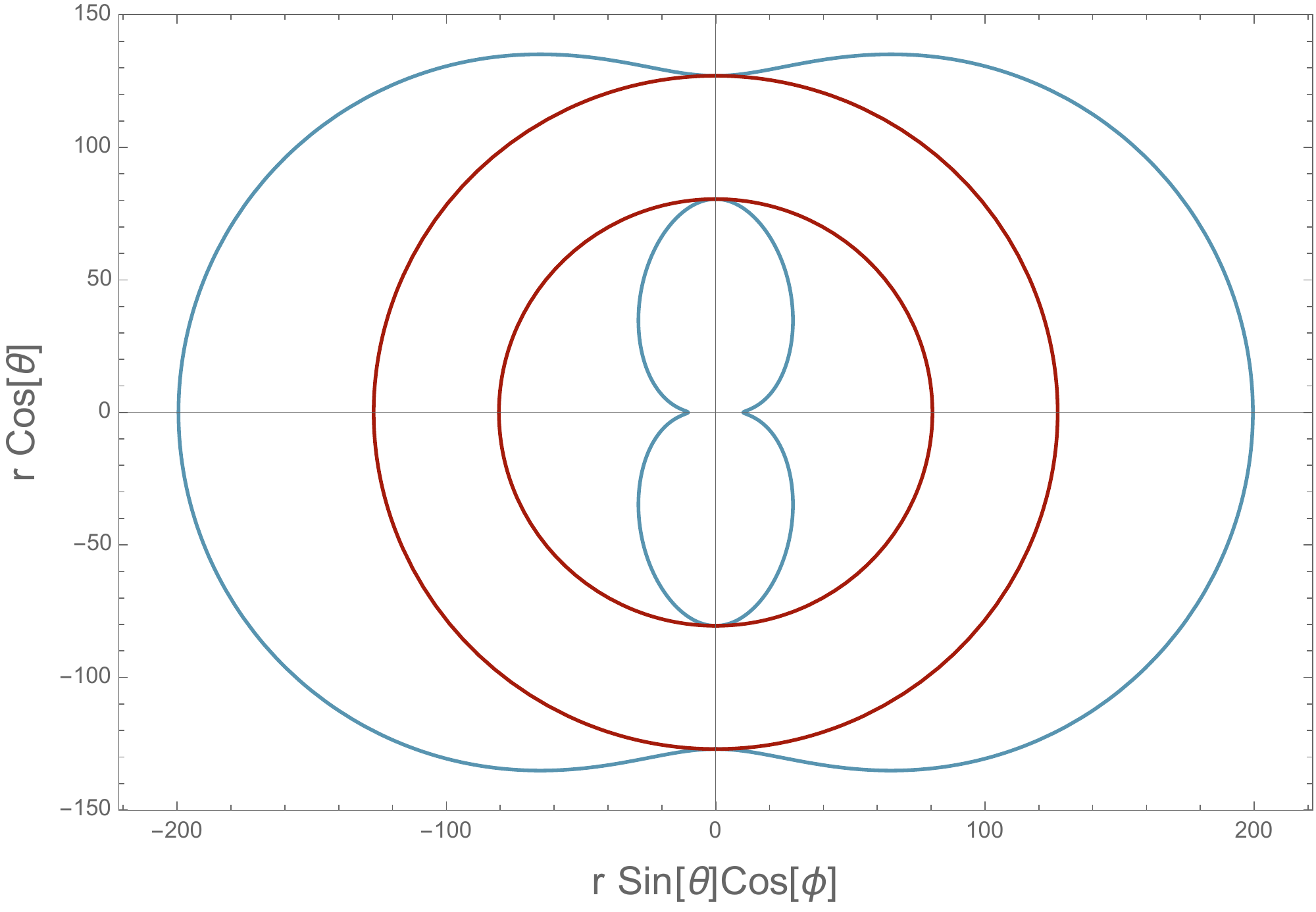}
\caption{Positions of the horizons (red) and stationary limit surfaces (blu), in the $r-\th$ plane. Here $\g=\d$, $m=100$, $a=95$ and $L=10$ (in Planck units), a configuration below the extremal case. As we approach the extremal case, the two horizons draw near and the hour-glass shape of the stationarity limit surfaces is enhanced.}
\label{fig:horizons}
\end{figure}

\paragraph{Ergosphere.} The stationary limit surfaces are defined by the vanishing norm of the time-like Killing vector, which as in Kerr, turns out to be given by 
\be\label{diciotto}
\tl\D - a^2\sin^2\th = 0.
\ee
Also this equation has at most 2 real positive solutions for $L\ll m$, which are $\th$-dependent, see Fig.~\ref{fig:horizons} for an example. 
The deformation induced by $L$ is similar to that of the horizons: the outer stationarity limit surface shrinks, while the internal one grows. The effect is stronger at the equator, and deforms the typical number-eight shape of the inner surface into an hour-glass shape. A similar deformation occurs with the rotating Bardeen black hole \cite{Ghosh:2015pra}. This has the consequence that the same Killing vector is time-like outside the outer stationary limit surfaces and inside the inner one, a fact that will play a important role later on. 

As for the horizons, we can solve perturbatively \Ref{diciotto} in the case $\g=\d$. 
The approximate locations of the stationary limit surfaces are
\be\label{rE}
r_\pm^{\rm E} + \eta_\pm L^2 + o(L^3),
\ee
where
\be
r_\pm^{\rm E} = m\pm \sqrt{m^2-a^2\cos^2\th}
\ee
are the usual locations in a Kerr black hole, and 
\be\label{etas}
\eta_\pm = -\f{2m}{(r^{\rm E}_\pm)^2}\f{(r^{\rm E}_\pm)^2+a^2\cos^2\th}{(r^{\rm E}_\pm)^2-a^2\cos^2\th}.
\ee
Expanding for $a\ll m$, we get
\begin{align}
& \eta_+ \simeq -\f{1}{2m}\left(1+\f{a^2}{m^2}\cos^2\th\right), \\
& \eta_- \simeq \f{8m^3}{a^4}\f1{\cos^4\th} +\f{1}{2m}\left(1+\f{a^2}{m^2}\cos^2\th \right).
\end{align}
We see that the approximation to the outer surface is valid for all $a\leq m$, whereas for the inner one it is only valid provided $m\geq a>\sqrt{m L}/\cos\th$.
Notice that at the equator the position of the outer stationarity limit surface matches the position of the event horizon of the non-rotating hole, as can be checked from \Ref{rE} and \Ref{etas}. The same relation exists between the Kerr and Schwarzschild metrics.

Among the consequences of a deformed ergosphere, there is Penrose's energy extraction process. The angular velocity of the horizon is
\be\label{Omega}
\Omega_H := -\f{g_{t\phi}}{g_{\phi\phi}} \bigg|_{r_+} = \f{a}{(r_+^{(0)})^2+a^2}(1-L^2\eps_+),
\ee
which is faster than Kerr for the same values of $m$ and $a$. As a consequence, the maximal extraction of angular momentum, $\d J = \d M/\Om_H$, decreases.

\begin{figure}[ht]
\centering
\includegraphics[width=0.4\textwidth]{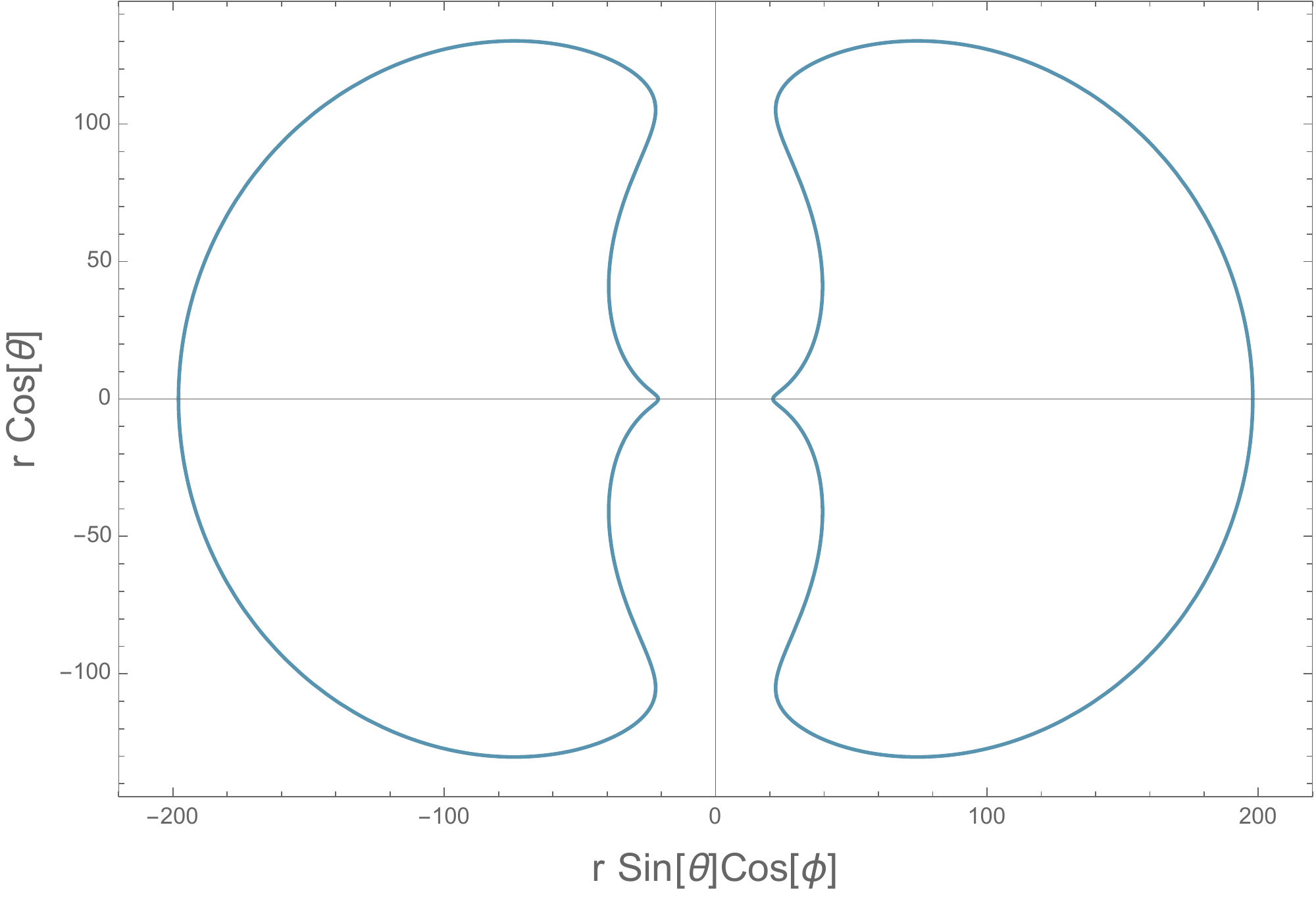}
\caption{The shape of the ergosphere in the ultra-extremal case. The horizons have disappeared, and the two stationary limit surfaces merge in the boundary of a 3-torus region.}
\label{FigErg}
\end{figure}

Of course, being regular the metric could make sense also beyond the extremal case, in which there are no more horizons, and the two stationary limit surfaces merge defining a compact surface with the topology of a 3-torus, see Fig.~\ref{FigErg}.
The astrophysical applications of such a metric are not clear, but given the absence of a singularity it is not obviously mandatory to peel away the excess angular momentum. It would then be worth investigating whether these configurations can be the end state of gravitational collapse in modified theories of gravity.

\begin{figure*}[!t]
\centering
\includegraphics[width=0.45\textwidth]{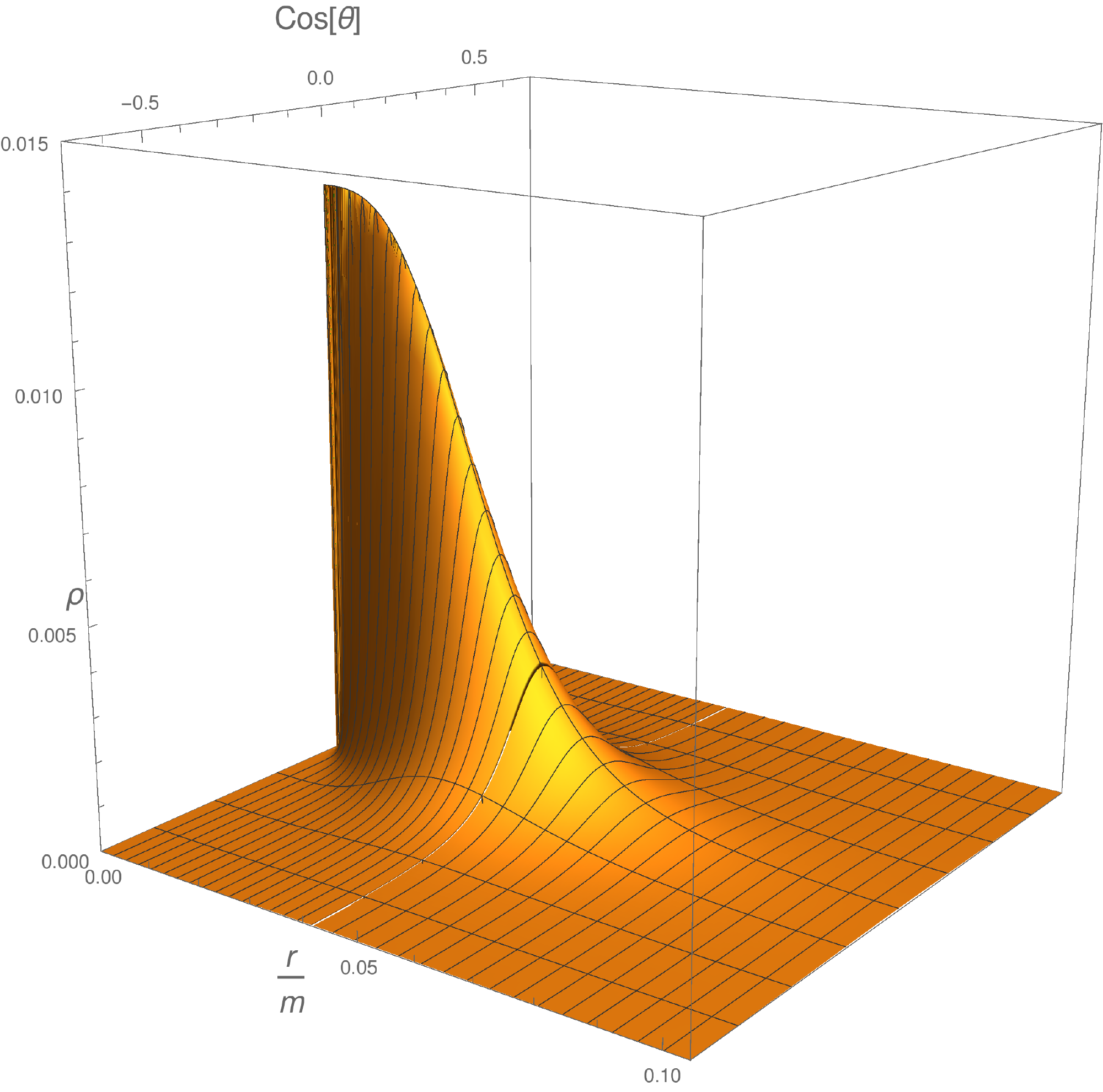}\qquad
\includegraphics[width=0.45\textwidth]{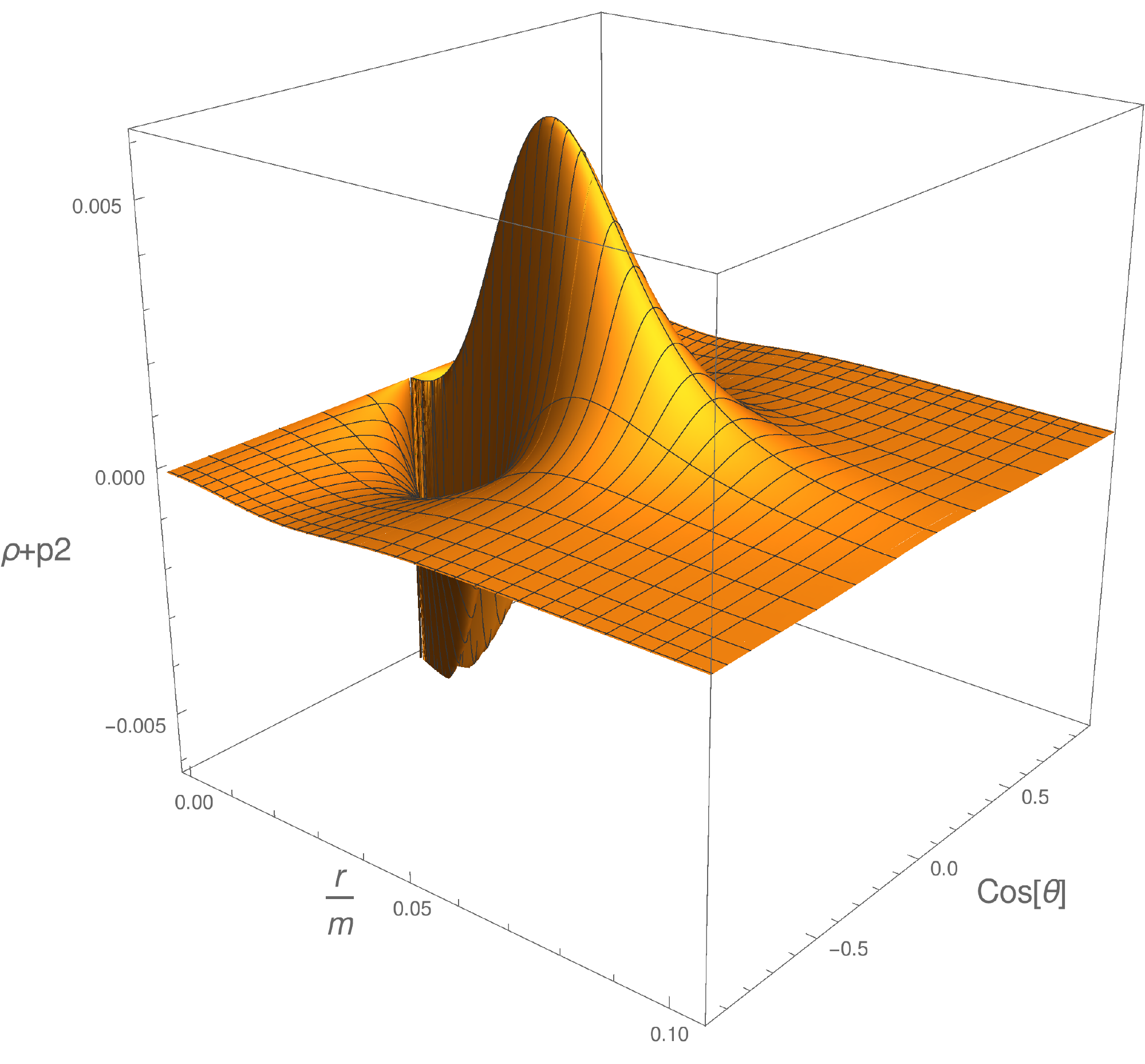}
\caption{Energy density and violation of the weak energy condition in one of the principal pressure, for the rotating Hayward spacetime with $m=500$, $a=100$, $L=3$ (in Planck units). The inner horizon is located at $r_-/m \simeq 0.04$ and event horizon at $r_+/m \simeq 2$. Both plots feature the same discontinuity at $r=0$ as for $\cal K$.}
\label{fig:wec}
\end{figure*}

\paragraph{Energy Conditions.}
We have already seen that parts of the Bambi-Modesto metric have a de Sitter-like behaviour, thus violating the strong energy condition.
It is well known that this type of non-singular black holes may also violate the weak energy condition. While this is not the case for Hayward's metric, it is so for the modified Hayward with a time delay in the centre \cite{DeLorenzo:2014pta}, and also for the Bambi-Modesto \cite{BambiModestoKerr}. The latter analysis is not straightforward, as the energy-momentum tensor is not diagonal in the initial coordinate system. But at least for the case $\g=\d$, it was explicitly shown in  \cite{BambiModestoKerr} that it can be diagonalised in two steps. First, projecting the components on the co-rotating tetrad of \cite{Bardeen72}, for which the energy-momentum tensor has a single non vanishing off-diagonal component. Then, finding a Lorentz transformation to complete the diagonalisation.\footnote{To be precise, in \cite{BambiModestoKerr} the authors consider a mixing of the tetrads that does not preserve the norms, and thus is not a Lorentz transformation. We believe however that their procedure amounts to a Lorentz transformation plus at most a permutation of the internal indices, and thus that their definitions of energy density and pressures coincide with ours.} 
This turns out to be a pure rotation in between the horizons, and a pure boost outside. As we show in Appendix A, the required boost is only admissible if in the projected indices $|T^{03}/(T^{00}+T^{33})|<1/2$, which puts (point-dependent) bounds on the parameters $(m,a,L)$. The bound can be satisfied  for a large range of realistic values of the parameters (that is, $L\ll a < m$), thus diagonalisation is possible in general, and the usual definition of the weak energy condition applies, $\rho\geq 0, \rho+p_i\geq 0$, with (in the internal indices of the tetrad) $T^{00}=\rho$ and $T^{ij}= p_i\d^{ij}$. In Fig.~\ref{fig:wec} we plot the weak energy condition obtained in this way, which confirms that the violation is confined to the principal pressures, and to a region around and inside the inner horizon. We also show that although the energy density is always positive, it is discontinuous at $r=0$, in agreement with $\cal K$.

When the above bound is violated, one may still try to find a different tetrad allowing to diagonalise the energy-momentum tensor, or at least to put it in one of the other canonical forms of \cite{segre1884sulla,hawking1973large}, in which the energy conditions can be explicitly given as a set of inequalities.

%--------------------------------------------------------------------------------------------------
\section{Introducing a time delay before the NJ algorithm}
%--------------------------------------------------------------------------------------------------

We now come to the main motivation for our work, that is the introduction of a non-trivial time delay between the regular center and spatial infinity. Indeed even a regular star, described for instance by the Schwarzschild interior solution with constant density, features a time delay -- dependent on the mass and the radius of the star -- between the core and spatial infinity, simply due to the fact that the star is not empty inside. 
Similarly, we expect that if a regular black hole represent a reasonable spacetime following on gravitational collapse of matter, it should allow for such a non-trivial time delay. Instead, an observer in the regular centre of the Hayward metric experiences no time delay with respect to an observer at infinity,
\be
(\delta t_{\infty} - \delta t_0)/\delta t_\infty = 1- \sqrt{|g_{00}(r=0)|} = 0.
\ee
In \cite{DeLorenzo:2014pta}, it was suggested to modify the Hayward metric in order to introduce a time delay requiring
\be\label{timedelay}
- g_{00}(r=0) = 1 - \alpha, \qquad \alpha\in[0,1).
\ee
The larger $\alpha$, the greater the time delay. This was achieved modifying $g_{00}$ with an extra factor, leading to a metric of the form \Ref{gensph} with $f(r)=F(r)G(r)$ and $h(r)=1/F(r)$, where $F(r)$ is of the Hayward type, and $G(r)$ a function subjected to a number of conditions. As a case study, a suitable function giving \Ref{timedelay} while preserving all other features of the Hayward metric (regularity and de Sitter core, horizon and asymptotic structures) was taken in the form of
\be\label{G3}
G(r) =  1 - \frac{\beta m\,\alpha}{\alpha\, r^3+\beta m}, \qquad G(0)=1-\a.
\ee

Following Bambi and Modesto's idea, we apply the Newman-Janis algorithm to this modified Hayward metric, using in particular the prescription \Ref{NJ3} for $G(r)$. We obtain a metric of the type \Ref{NJg}, with $\tl f(r)=\tl F(r)\tl G(r)$ and $\tl h(r)=1/\tl F(r)$, describing a rotating black hole with mass $m$ and angular momentum $a$, as desired. 
Furthermore, since $G(r)$ has no zeroes, the location of the horizons and stationarity limit surfaces are unchanged.
However, it turns out to have a curvature singularity in the centre, unlike the seed metric. 

To show this, let us first consider the case $\g=\d$, in which both $M(r)$ and $G(r)$ receive no deformations from the NJ algorithm.
A power series expansion near $\th=\pi/2$ of the Kretschmann scalar evaluated at the origin gives
\be\label{KGdiv}
{\cal K}|_{r=0} \sim \f{c_1}{(\th-\f{\pi}{2})^6} + \f{c_2}{(\th-\f{\pi}{2})^4} +\f{c_3}{(\th-\f{\pi}{2})^4}+ O(1),
\ee
where the coefficients of the diverging terms are given by
\be\nonumber
\begin{split}
c_1 &=\f{8}{a^4}\big(\a+2\,\sqrt{1-\a}-2\big), \\
c_2 &=\f{4}{a^4}\big(80-80\sqrt{1-\a}-84\,\a  \\
&\qquad\qquad\qquad + 44\,\a\sqrt{1-\a}  + 11 \a^2\big), \\
c_3 &=\f{8}{15 \, a^4}\big(404\,(1-\sqrt{1-\a})\\
&\qquad\quad +(220\,\sqrt{1-\a}-422)\,\a +55\,\a^2\big)\,.
\end{split}
\ee
The three coefficients have to vanish simultaneously for $\cal K$ to be finite, and by inspection, it is easy to see that the only real solution is $\a =0$. That is, $G(0)=1$ and no time delay is allowed. 

One may then ask whether a different $G(r)$ would work. Notice that a different function with a priori more powers of $r$ appearing means also more freedom than \Ref{NJ1} and \Ref{NJ3} in the complexification procedure. Let us assume that whatever the complexification procedure is, it does not introduce a dependence on $\th$. 
Then, taking an arbitrary $\tl G(r)$, and keeping $\g=\d$ so that $M(r)$ is not modified, the power series expansion near the equator has the same form \Ref{KGdiv}, where the coefficients
are now functions of $\tl{G}(0)$, $\tl{G}'(0)$ and $\tl{G}''(0)$. Their explicit form is too long to be reported here, but details are given in Appendix B. The key point is that the system $c_1=c_2=c_3=0$ admits again a unique real solution given by $\tl{G}(0)=1$. 

In the case $\g\neq \d$ an explicit evaluation of the power series is not possible, because $\g$ and $\d$ appear as powers of $r$. However, the proof of existence of divergences can be extended by a simple argument that exploits the matching of the metric and its first derivatives with the $\g= \d$ case. Details are again reported to the Appendix C. The only case that we leave open is an arbitrary seed function or complexification procedures such that $\tl G=\tl G(r,\th)$. In this case, we are not able to explicitly evaluate the analytic behaviour of the Kretschmann invariant at the core, and the argument used in the Appendix does not apply since the metrics and first derivatives do not match.

We conclude that, with the possible exception of a $\th-$dependent time delay function, the NJ algorithm applied to the modified Hayward metric with a time delay in the centre gives a rotating singular black hole.

%--------------------------------------------------------------------------------------------------
\section{Introducing time delay after the NJ algorithm}\label{sec:timedelay}
%--------------------------------------------------------------------------------------------------

Since we could not obtain the desired metric applying the NJ algorithm to the non-rotating one, the alternative solution is to introduce a time delay directly at the level of the rotating Bambi-Modesto metric.
To that end, let us restrict ourselves to the special case $\g=\d$, where we can use the metric in the Boyer-Lindquist form, as in equation~\Ref{BM1}.
As discussed above in the analysis of the ergosphere, the deformation of the inner stationarity limit surface guarantees that time in the centre is still measured by $g_{00}(0)$, 
so it is possible to proceed as in the spherically symmetric case, introducing a function $G(r)$ modifying $g_{00}(0)$.
However, there are now different natural choices, since the non-diagonal structure of the metric mixes $t$ and $\phi$:
{\renewcommand{\theenumi}{ (\roman{enumi})}
\renewcommand{\labelenumi}{\theenumi}
\begin{enumerate}
\item $g_{00}\mapsto G(r) \,g_{00}$, modifying only the $dt^2$ term in the metric. 
\item $dt \mapsto \sqrt{G(r)} \,dt$, modifying both the $dt^2$ and $dtd\phi$ terms. 
\item $dt \mapsto \sqrt{G(r)} \,dt$ and $d\phi \mapsto \sqrt{G(r)} \,d\phi$, modifying also the $d\phi^2$ term. 
\end{enumerate}
}

All three cases lead to the same time delay in the centre, \Ref{timedelay}, and do not modify neither the position of the horizons, nor of the stationarity limit surfaces. On the other hand, they can be physically characterised looking at the frame dragging they induce. This is given by
\be
\om = -\f{g_{t\phi}}{g_{\phi\phi}}
\ee
which is unchanged in cases (i) and (iii), while it peaks up a factor $\sqrt{G(r)}$ in case (ii). 
Hence, also the angular velocity of the horizon $\Om_H$ in eq.~\Ref{Omega}, and thus the maximal energy extraction via the Penrose process, are modified. As $G(r)<1$, the effect of introducing a time delay in this way is to reduce the angular velocity. Case (iii) can then be distinguished from (i) looking at the horizon area. We also point out that while case (i) may appear the simplest, it is harder to find a co-rotating tetrad for it, as the family proposed in \cite{Bardeen72} does not adapt in a simple way.

Notice that the metrics obtained in this way are not of type \Ref{NJg}. In particular, the time-delay function appears in at most three components, and in a less involved way than when introduced through the NJ algorithm. The power series expansion of $\cal K$ still has the structure \Ref{KGdiv}, but the coefficients change significantly. Their explicit expressions in terms of $G(r)$ and its derivatives are reported in Appendix B. As we show there, it is possible to have $c_1=c_2=c_3=0$ with $G(0)\neq 1$, provided that the first two derivatives of $G(r)$ vanish in the limit $r\to 0$.
Due to the possible discontinuity of $\cal K$, we also have to impose the finiteness of $\cal K$ as we approach the centre along the equatorial plane. A power series expansion in $r$ at the equator gives
\be\label{KGdivEq}
{\cal K}|_{\th=\pi/2} \sim \sum_{n=1}^6 \f{d_{(n)}}{r^n} + O(1),
\ee
with the coefficients of the six divergent terms reported in Appendix B. As shown there, they can be all put to zero as long as the first three derivatives of $G(r)$ vanish in $r=0$. Hence, a time delay is after all compatible with rotation and absence of singularity, if one introduces it in this way.

Interestingly, the first finite term of the two expansions can never be made to match. Indeed, the first finite term of \Ref{KGdiv} is always zero, whereas that of \Ref{KGdivEq} is a function of $G(0)$ and its fourth derivative $G^{(4)}(0)$, quadratic in the latter but with no real zeros. 
Hence, the discontinuity of the Kretschmann invariant can not be resolved with a time delay function. 
The Hayward value $\lim_{r\mapsto 0}{\cal K}|_{\th=\pi/2}=24/L^4$ is again recovered for $G^{(4)}(0)=0$. 

Although the explicit form of the coefficients differ, the conclusions on the finiteness conditions and the discontinuity hold for all three cases (i-iii).

The finiteness requirements discussed above impose the vanishing of one derivative more that in the static case \cite{DeLorenzo:2014pta}, so we can not use the profile \Ref{G3}. Of course, there is an infinite freedom in choosing a function with the required vanishing derivatives and unity asymptote at spatial infinity. In the static case it was also useful to require monotonicity in $r$, as it was a necessary -- but not sufficient -- condition to preserve the weak energy condition \cite{ dymnikova2002cosmological, DeLorenzo:2014pta}. Preserving the weak energy condition in the rotating case appears far from obvious (see e.g. the analysis of \cite{NevesSaa}), but we still appreciate the monotonicity condition, so that $0<G(r)<1$ and the main large scale features are still controlled by $\tl F(r)$, and we hope that it also helps minimising the violations. Accordingly with the above conditions, we choose a new time-delay function given by
\be\label{Gexp}
G(r)= 1- \a +\a \exp\left[-\f{\beta m}{\a r^3}\right]
\ee
(extended by continuity at $r=0$).
Here $\a$ is the same time delay parameter as in \Ref{timedelay}, and $\b$ a coefficient parametrising the $1/r^3$ large scale behaviour.

\begin{figure}[ht]
\centering
\includegraphics[width=0.45\textwidth]{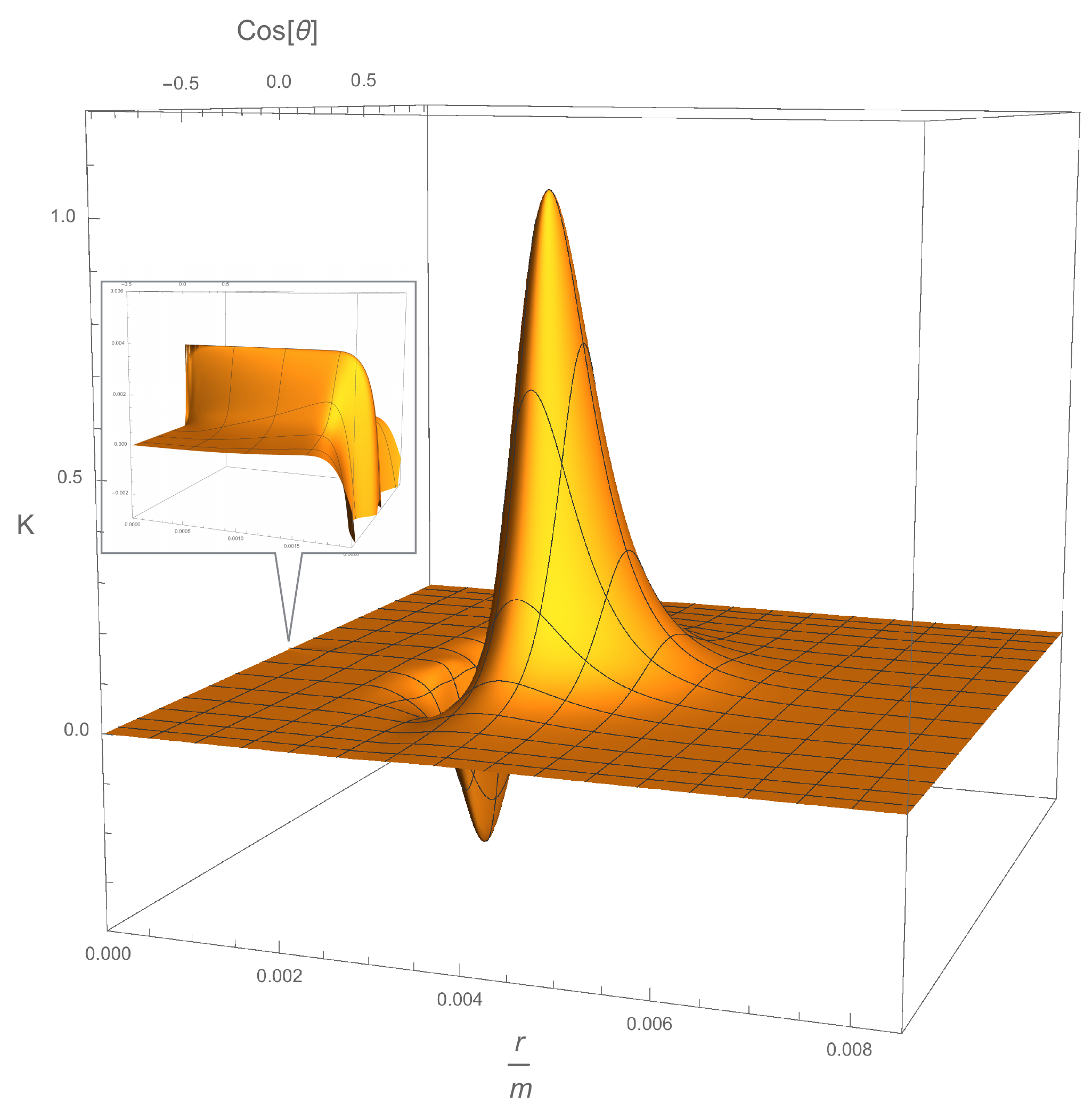}
\caption{The Kretschmann invariant ${\cal K}$ for the rotating Hayward metric with a time delay given by \Ref{Gexp}, with (in Planck units) $m=5000$,  $a=40$, $L=9$, $\a=0.8$ and $\b=41/(10\pi)$. The (discontinuous) peak at $r=0$ shown in the zoomed panel on the left is now dwarfed by the absolute maximum that develops just outside the inner horizon, located at  $r_-/m \simeq 0.004$. Notice also the negative values between the two maxima. Both features are also present in the non-rotating metric with time delay at the centre.}
\label{fig:K}
\end{figure}

The numerical evaluation of the Kretschmann scalar for \Ref{Gexp} in case (ii), is shown in Fig.~\ref{fig:K}.
It is finite everywhere, as expected. Notice that it has a non-monotonic behaviour, and a maximum displaced from the centre to just outside the inner horizon. The same two effects occur in the non-rotating case, and it is a clear feature of having $|g_{00}|<1$.
As we change the value of the parameters, the value of the maximum can exceed 1 in Planck units. Hence, again as in the static case, there will be an upper bound on the allowed time delay imposed by the requirement of a sub-Planckian curvature.

\begin{figure*}[t!]
\centering
\includegraphics[width=0.47\textwidth]{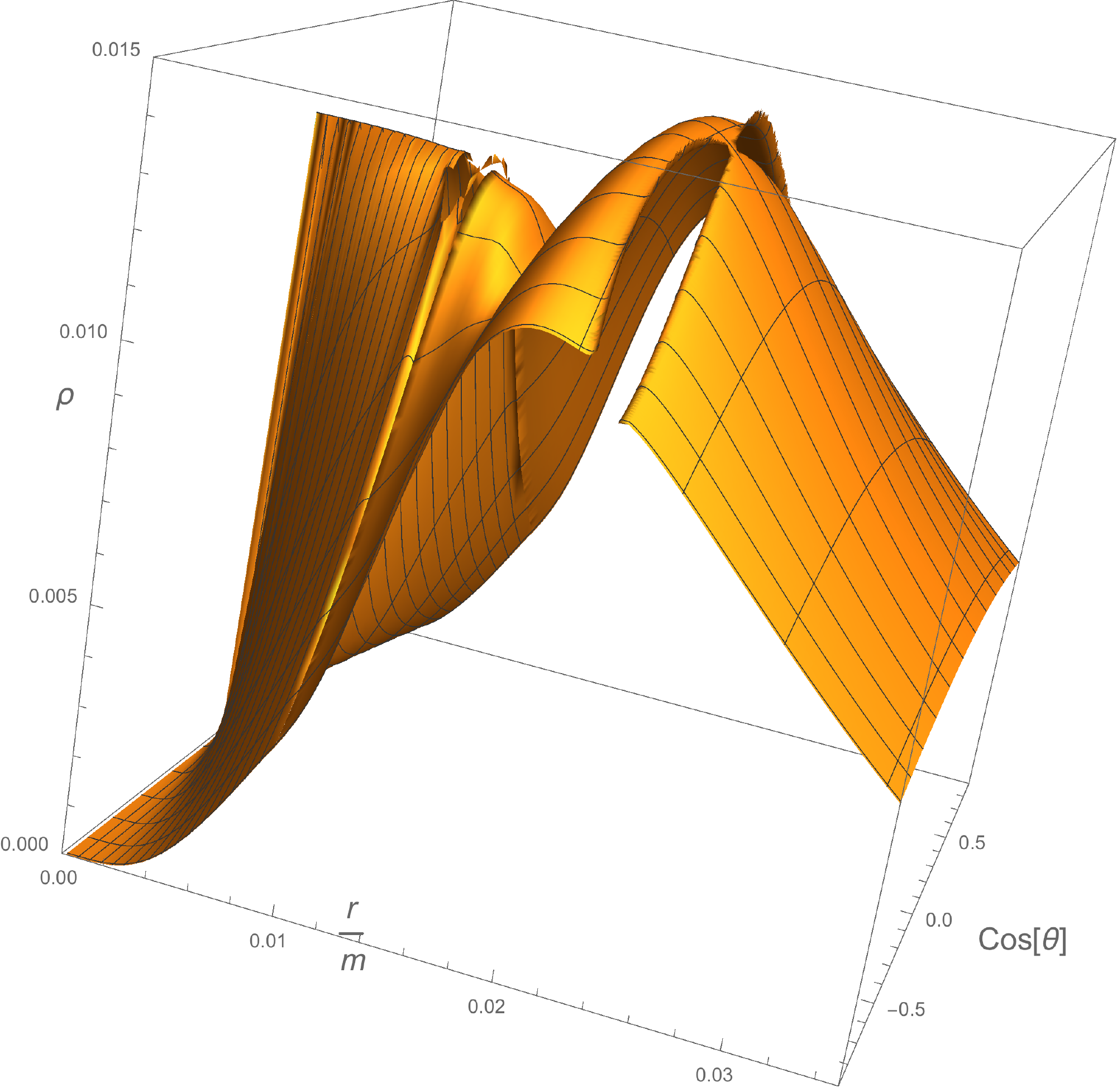}\qquad
\includegraphics[width=0.47\textwidth]{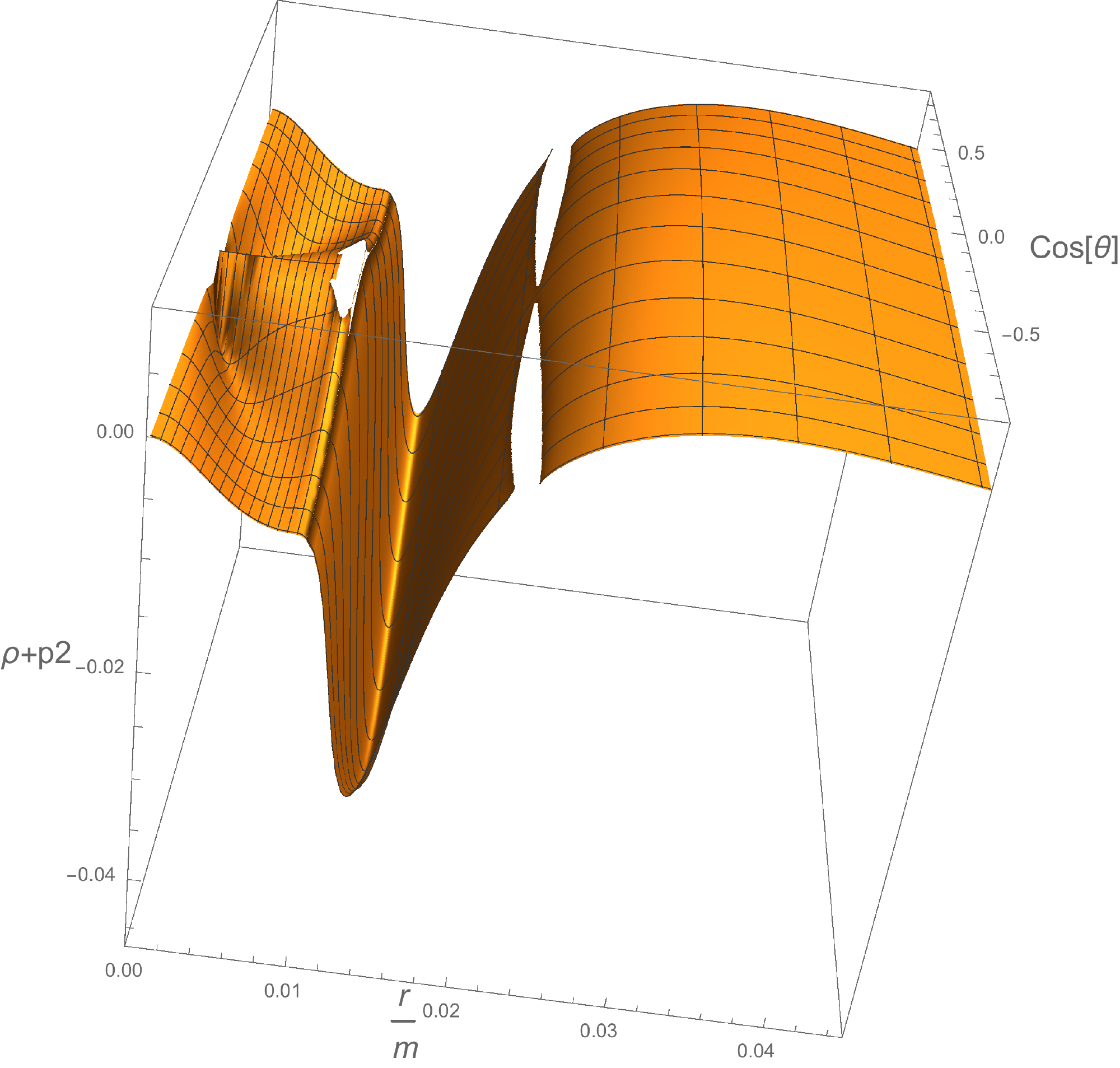}
\includegraphics[width=0.47\textwidth]{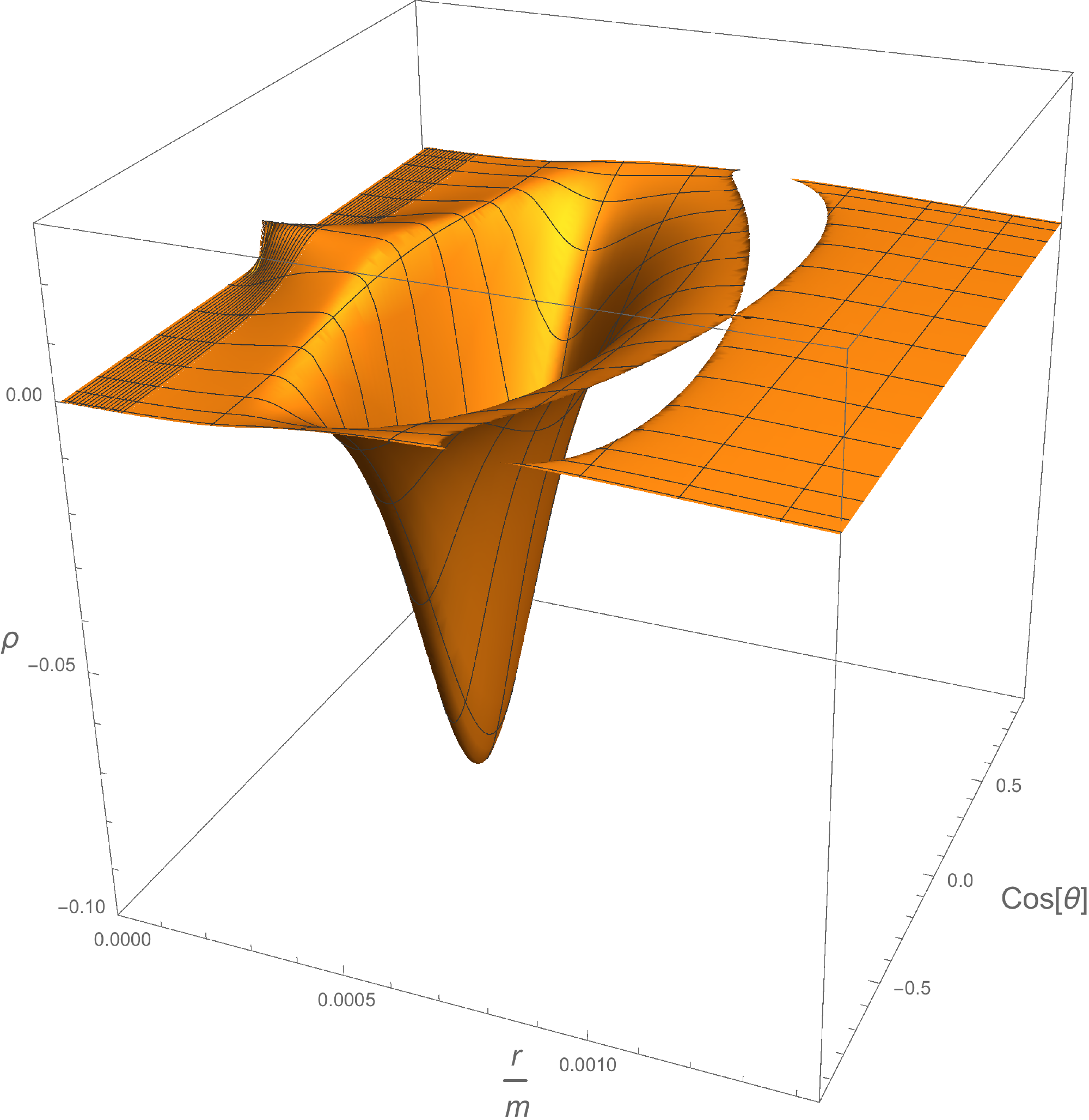}\qquad
\includegraphics[width=0.47\textwidth]{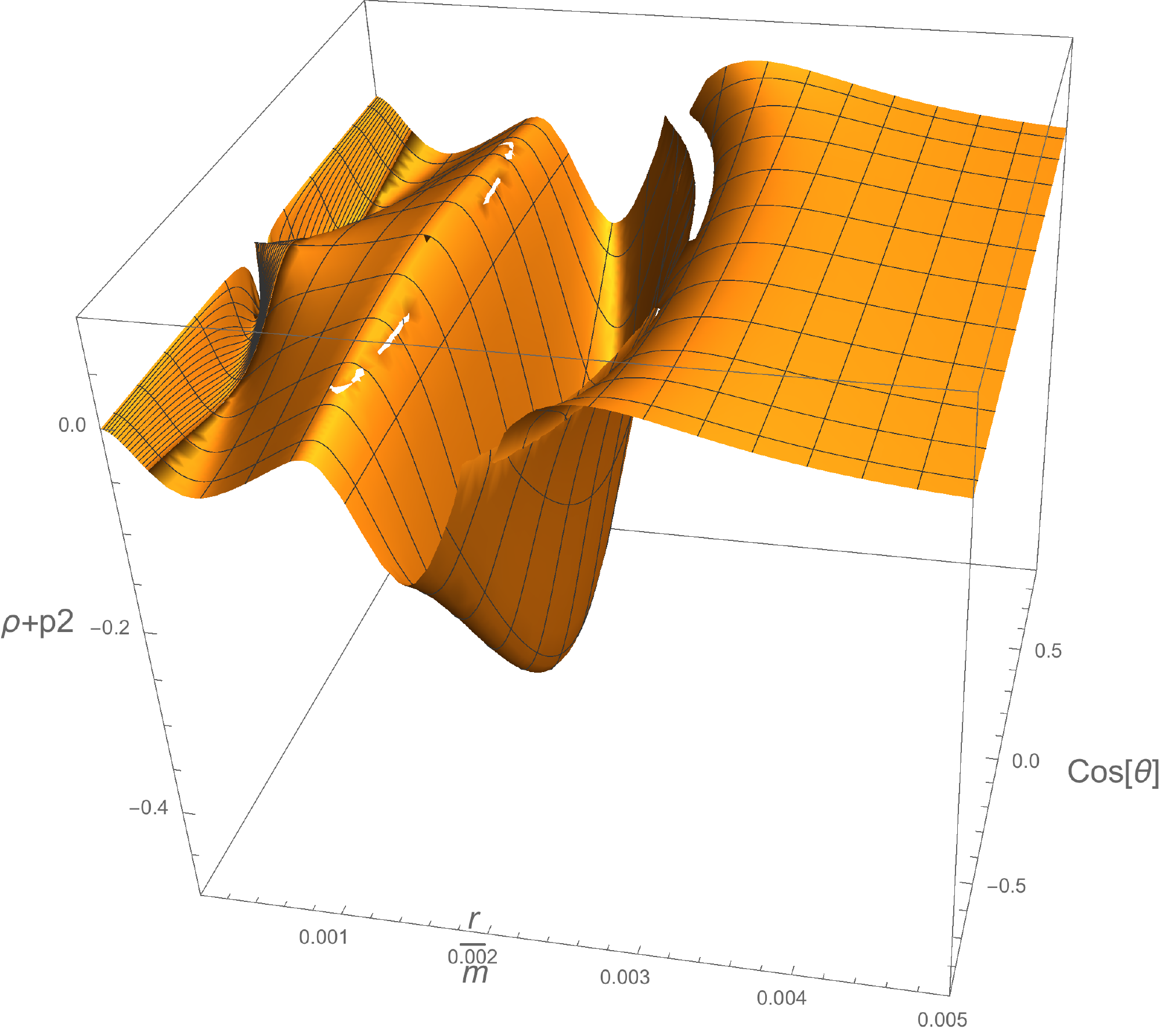}
\caption{\emph{Top panels:} The energy density and a representative $\rho+p_2$ weak energy condition for the rotating metric with a time delay given by \Ref{Gexp}, case (ii), with (in Planck units) $m=500$, $a=5$, $L=3$, $\a=0.9$ and $\b=41/(10\pi)$. The weak energy condition is violated only in the principal pressures, in a region just outside the inner horizon located at $r_-/m \simeq 0.009$. 
The event horizon is at $r_+/m \simeq 2$, and the energy momentum tensor goes smoothly to zero well before. The voids correspond to regions where the energy-momentum tensor can not be diagonalised with the co-rotating tetrad. Some numerical noise is present at the boundary of the void regions.
\emph{Bottom panels:} Negative energy density appearing for $m=10^5$, $a=50$, $L=4$, $\a=0.8$ and $\b=41/(10\pi)$, with inner horizon at $r_-/m \simeq 0.00014$ and event horizon at $r_+/m \simeq 2$.}
\label{fig:Negrho}
\end{figure*}

The energy-momentum tensor defined by this metric is not diagonal. To study the weak energy condition, we proceed as in \cite{BambiModestoKerr}, and use the co-rotating tetrad of \cite{Bardeen72} to simplify the analysis. However, this tetrad only works for cases (ii) and (iii), to which we restrict attention. The numerical analysis, of which we report in Fig.~\ref{fig:Negrho} two typical cases, shows that the energy-momentum tensor goes smoothly to zero at large scales, but the weak energy condition is again violated in a zone inside and around the inner horizon. The energy density profile also changes, in agreement with the non-monotonic behaviour of $\cal K$, and discontinuities are present at $r=0$.
For some values of the parameters, the weak energy condition can also be violated by regions of negative energies, again only in the vicinity of the inner horizon, see bottom panel of Fig.~\ref{fig:Negrho}. 

Notice also from the figure the presence of void zones. These correspond to places where the energy-momentum tensor can not be diagonalised starting from the co-rotating tetrad. Indeed, the energy-momentum tensor  projected on the co-rotating tetrad has two non-vanishing off-diagonal components, as opposed to the single one of the Bambi-Modesto metric. As a consequence, diagonalising it requires a boost also in between the horizons, with the corresponding additional restriction 
$|T^{12}/(T^{11}+T^{22})|<1/2$. While in the case without time delay the point-dependence was rather weak, having now one more condition imposes strong limitations on the procedure: for the same generic range of parameters we always find some finite region, either in between the horizons or outside, 
 where the energy-momentum tensor is not diagonalisable with the co-rotating tetrad. This shows up in the void zones in the numerical plots. 

Let us finally comment on the parameter $\b$. A $1/r^3$ decaying factor is predicted by the 1-loop corrections in perturbative quantum gravity, for both the Schwarzschild and the Kerr metrics \cite{DonoghueKerr}. This factor is missing in the static Hayward case, whose corrections begin at $1/r^4$, thus the function $G(r)$ also allows to match the 1-loop correction, suitably fixing the parameter $\b$ with the calculations of \cite{DonoghueKerr}.
The Bambi-Modesto metric already has corrections in $1/r^3$ (for instance $g_{00}$ has $m a^2 \cos^2\th/r^3$), but one still needs to add a further correction if the result of \cite{DonoghueKerr} is to be reproduced. There is actually one difficulty in doing so explicitly: the perturbative quantum gravity calculations are performed in the harmonic gauge, and finding it in our case requires solving a differential equation that can be done only numerically. For this reason, we refrain from attempting such a matching here, and we leave the question open for future work.

%--------------------------------------------------------------------------------------------------
%--------------------------------------------------------------------------------------------------
\section{Conclusions}
%--------------------------------------------------------------------------------------------------
%--------------------------------------------------------------------------------------------------

Non-singular black holes are interesting low-energy models of quantum gravity theories, with applications to the understanding the final stages of Hawking evaporation and the information loss paradox, which are among the main motivations for introducing them, see e.g. \cite{hayward2006formation,Visser:2009pw,Frolov:BHclosed,Perez:2014xca}. Following on the analysis of \cite{Bianchi:2014bma}, it is possible to see \cite{SmerlakSeminar,de2014investigating} that simple evaporation models based on the spherically symmetric Hayward metric do not satisfy energy conservation laws or existing purification bounds such as those of \cite{Carlitz:1986ng}. Hence, more work is needed towards the construction of a physically acceptable metric, and we hope that the results of this paper are a small step in that direction.

To summarize our results, we investigated the Bambi-Modesto metric for a rotating, regular black hole. We exhibited the role of the quantum gravity parameter in approaching the horizons, decreasing the maximal angular momentum of the extremal case, and deforming the ergosphere. We then showed that the Newman-Janis algorithm produces a singularity in the curvature when applied to a non-singular black hole with a time delay in the centre. Finally, we showed how it is possible to introduce the time delay directly at the level of the rotating metric. The procedure is not unique, and we characterised the different possibilities in terms of the area and velocity of the horizon. 

Regularising a black hole metric of course also means taking seriously the whole spacetime region inside the event horizon, dealing with the inner Cauchy horizon and with the possible existence of closed time-like curves. Using \Ref{BM1} or the modified versions with a time delay, it is easy to see that there are no closed causal curves with $r$ always positive. In the Kerr solution, closed time-like curves appear for path crossing to the $r<0$ maximal extension of the spacetime. It would indeed be interesting to study the geodesics equation of the non-singular rotating metrics, to understand the consequences of the discontinuity in the curvature invariants and whether extensions of the spacetime may arise in spite of the absence of singularities. We leave this question open for future work.

%--------------------------------------------------------------------------------------------------
%--------------------------------------------------------------------------------------------------
\subsection*{Acknowledgments}
%--------------------------------------------------------------------------------------------------
%--------------------------------------------------------------------------------------------------
We thank Pietro Don\`a, Thibaut Josset and Carlo Rovelli for useful discussions.

\AppSep
\appendix
%--------------------------------------------------------------------------------------------------
\subsection*{A. Diagonalising $T^{\m\n}$}
%--------------------------------------------------------------------------------------------------
The co-rotating tetrad $e^I_\mu$ of \cite{Bardeen72} in the Bambi-Modesto metric has norms given by
\be\label{norms}
g^{\mu\nu}e^I_\mu e^J_\nu =\begin{cases} (-{\rm Sign}(\tl \Delta),{\rm Sign}(\tl \Delta),1,1) & {\rm if\;} I=J\\
0 & {\rm if\;} I\neq J
\end{cases}
\ee
The corresponding Einstein tensor defines a non-diagonal energy-momentum tensor of the form
\be
T^{IJ} = \left(\begin{array}{cccc} i & 0 & 0 & j \\ 0 & k & 0 & 0 \\
0 & 0 & l & 0\\j & 0 & 0 & n   
\end{array}\right).
\ee
We then ask whether it is possible to diagonalise this tensor with a Lorentz transformation. To that end, we can concentrate on the 2-by-2 block with $I=0,3$. The transformation will have to be either a simple rotation, when ${\rm Sign}(\tl \Delta)=-1$, between the event and the Cauchy horizons, or a boost, when ${\rm Sign}(\tl \Delta)=+1$ in the rest of the spacetime. Consider first the latter case. The most general $(1+1)$ boost reads
\be
\Lambda^I{}_{J}=\left(\begin{array}{cc} \cosh \eta & \sinh \eta \\ \sinh \eta & \cosh \eta \\  
\end{array}\right)\,.
\ee
The condition $D^{03}=0$, where $D^{IJ}=\Lambda^I{}_{K} T^{KL}\Lambda^J{}_{L}$, reduces to
\be\label{taneta}
\tanh \eta = -\f{2j}{i+m}
\ee
and implies 
\be\label{cond}
\left|-\f{2j}{i+m}\right|<1
\ee
in order for the transformation to be valid, which in turn imposes conditions on the parameters of the metric. In between the horizons, since the transformation is a rotation, this problem does not arise because condition \Ref{taneta} is replaced by  
\be
\tan \eta = -\f{2j}{m-i}
\ee
which can be always satisfied.
When condition \Ref{cond} is not satisfied -- notice that this may also happen locally, as the coefficients in \Ref{cond} are spacetime functions --, it is not possible to diagonalize the energy momentum tensor starting from the co-rotating tetrad. Therefore, we need to start from scratch to find the good orthonormal basis. This is generally a not easy task and we do not address this problem here.

When a time delay is introduced, for cases (ii) and (iii) is still possible to define the co-rotating tetrad with the norms given in eq.~\Ref{norms}. In these cases, however, the transformed energy-momentum tensor $T^{IJ}$ has two non-null components out of the diagonal, namely $T^{03}$ and $T^{12}$. To diagonalize it, therefore, we need now to combine a boost and a rotation for the two different 2-by-2 blocks according to the sign of $\tl \Delta$. This introduces the same condition \Ref{cond} when ${\rm Sign}(\tl \Delta)=-1$, together with a similar one for the $1-2$ block of the tensor in the opposite case.
The additional condition imposes strong limitations on the procedure, so that for the same range of parameters we now typically have finite regions where the energy-momentum tensor is not diagonalisable with the co-rotating tetrad. This shows up in the void zones in the numerical plots. The void regions can be either in between the horizons or outside, depending on the value of the parameters.

%--------------------------------------------------------------------------------------------------
\subsection*{B. Coefficients of the $\cal K$ expansion}
%--------------------------------------------------------------------------------------------------
We report in this Appendix various coefficients of power series used in the main text. For the expansion \Ref{KGdiv}, we have
\be\nonumber
c_1=\f{1}{a^4\,G(0)}\Big[ 8G(0)\Big(2 \sqrt{G(0)} - 1 - G(0)\Big) + a^2 G'(0)^2 \Big],
\ee
and the other two, $c_2$ and $c_3$, are functions of $G(0)$ and its derivatives up to the second order, too long to be written here. However, once the solution for $c_1=c_2=0$ is plugged into the equation $c_3=0$, one gets the following relatively simple equation for $G(0)$,
\be\label{c3}
\begin{split}
&G(0) \big( 100\,\sqrt{G(0)} - 21 \big)\\
&\quad - G(0)^2 \big(51\, G(0) - 44\,\sqrt{G(0)} - 86 \big)\\
&\qquad+ \Gamma\; \big(1-G(0)\big)=0,
\end{split}
\ee
where the function $\Gamma$ is defined by
\be\nonumber
\Gamma^2:= 32\, G(0)^2\,(\sqrt{G(0)}-1)^2(55\,G(0)+10\sqrt{G(0)}-9)\;.
\ee
The only positive solution of eq.~\Ref{c3} is $G(0)=1$.

On the other hand, when the time delay is introduced after the NJ procedure, non-trivial solutions exist.
The expansion still has the structure \Ref{KGdiv}, but with different coefficients. Both cases (i) and (ii) give the same result,
\be\nonumber
\begin{split}
& c_1 = 2\f{G'(0){}^2}{a^2G^2(0)}, \\
& c_2 = \f{(G'(0){}^2-2G(0) G''(0))^2}{4G(0)^4}\\
& c_3 = \f23 c_2 - 4 \f{G'(0){}^2}{15 a^2 G(0)^2}
\end{split}
\ee
which can be all made to vanish for $G'(0)=G''(0)=0$, while keeping an arbitrary $G(0)\neq 0$.
While the coefficients are slightly different in case (iii), their vanishing leads to the same condition.

Finally, the coefficients of the equatorial expansion
\Ref{KGdivEq} are much longer. Explicitly for case (ii), the first three read
\be\nonumber
\begin{split}
& d_6 = \f{2a^4 G'(0)^2}{G(0)^2}, \\
& d_5 = \f{a^4 G'(0)}{G(0)^3}(2G(0)G''(0)-3G'(0)^2), \\
& d_4 = \f{a^2}{4G(0)^4L^2}\bigg[16 a^2G(0)^2G'(0)^2 \\
&\quad + L^2 \Big[4a^2G(0)^2G''(0)^2 + G'(0)^2\Big(13a^2G'(0)^2 \\
&\quad -16a^2G(0)G''(0)+8G(0)^2\Big)\Big]\bigg].
\end{split}
\ee
These vanish iff the first two derivatives vanish. With this condition, $d_3\equiv 0$ and
\[
d_2 = \f{G^{(3)}(0)^2}{2G(0)^2},
\]
which gives the additional condition $G^{(3)}(0)=0$ for the third derivative.
Finally, the three conditions all together make $d_1$ vanish.

%--------------------------------------------------------------------------------------------------
\subsection*{C. Divergence for $\g\neq \d$}
%--------------------------------------------------------------------------------------------------
In Section~\ref{sec:timedelay} and the previous Appendix, we showed that introducing a time delay in the rotating case by applying the NJ algorithm to the metric of \cite{DeLorenzo:2014pta} unavoidably leads to a divergent $\cal K$, for any function $G(r)$ and for a complexification such that $M(r)$ and $G(r)$ are unchanged, as choosing $\g=\d$ in the Bambi-Modesto prescription.
When $\g\neq \d$, the dependence on $r$ and $\th$ is too complicated to be handled explicitly, and we can not derive a power series expansion near zero. However, the divergence can be established with an indirect argument. In fact, notice that at the equator, where the divergence lurks, the metric for $\g\neq \d$ coincides with the metric for $\g=\d$. Furthermore, also the first derivatives coincide, and most of the second derivatives. The only terms that differ are second derivatives in $\theta$ of three metric components. Explicitly,
\begin{equation} \nonumber
\partial^2_\theta(g-\bar{g}) \big|_{\theta=\frac{\pi}{2}}=  (\gamma - \delta) \frac{8 a^2 m^2L^2 G(r)}{(4mL^2 + r^3)^2}\times
\begin{dcases*}
-1 & $(uu)$\\
a & $(u\phi)$\\
-a^2 & $(\phi\phi)$\\
\end{dcases*}
\end{equation}\\
where $g$ is the metric for $\g \neq\d$, while $\bar{g}$ is the one with $\g = \d$.

Hence, the Riemann tensors evaluated at the equator only differ in the terms of type $R^{\m}{}_{\th\th \n}$, and the difference of the Kretschmann invariants is
\be\label{Kdiff}
({\cal K} - \bar{\cal K})\big|_{\theta=\f\pi2} = 8 R^\m{}_{\th\th\n} A_{\m}{}^{\n} + 2 A^{\m\n} A_{\m\n},
\ee
where
\be
A^\m{}_\n := \f12\,\bar{g}^{\m\l}\,\p_\th^2(g_{\l\n}-\bar{g}_{\l\n})\big|_{\theta=\f\pi2}.
\ee\\
The latter quantity is zero except for the $u$ and $\phi$  components and, more importantly, it is finite. Eq.~\Ref{Kdiff}, therefore, tells us that the divergence of one implies the divergence of the other.

\AppSep

\providecommand{\href}[2]{#2}\begingroup\raggedright\endgroup

\end{document}